\documentclass[aps,prd,groupedaddress,nofootinbib]{revtex4-2}

\pdfoutput=1

\usepackage{float}
\usepackage{amsmath}
\usepackage{color}
\usepackage{graphicx}
\usepackage[dvipsnames]{xcolor}
\usepackage{url}
\usepackage{epsfig}
\usepackage[T1]{fontenc}
\usepackage{multirow}
\usepackage{physics} 
\usepackage{booktabs} 
\usepackage{array} 
\usepackage{paralist} 
\usepackage{grffile}
\usepackage{verbatim} 
\usepackage{subfig} 
\usepackage{amsmath,amsthm,amssymb,bm,amsfonts}
\usepackage{slashed}
\usepackage[utf8]{inputenc}
\usepackage{hyperref}
\allowdisplaybreaks[1]

\usepackage{natbib}
\usepackage{color}
\usepackage[normalem]{ulem}

\def\beg{\begin{equation}}
\def\eeg{\end{equation}}
\def\bea{\begin{eqnarray}}
\def\eea{\end{eqnarray}}

\usepackage{multirow}
\usepackage[title]{appendix}

\newcommand{\slv}{\raise.15ex\hbox{$/$}\kern-.53em\hbox{$v$}}
\newcommand{\slnbar}{\raise.15ex\hbox{$/$}\kern-.53em\hbox{$\bar{n}$}}
\newcommand{\slF}{\raise.15ex\hbox{$/$}\kern-.53em\hbox{$F$}}
\newcommand{\sllbar}{\raise.15ex\hbox{$/$}\kern-.40em\hbox{$\bar{l}$}}
\newcommand{\slh}{\raise.15ex\hbox{$/$}\kern-.40em\hbox{$h$}}
\newcommand{\slP}{\raise.15ex\hbox{$/$}\kern-.53em\hbox{$P$}}
\newcommand{\slR}{\raise.15ex\hbox{$/$}\kern-.53em\hbox{$R$}}
\newcommand{\slz}{\raise.15ex\hbox{$/$}\kern-.53em\hbox{$Z$}}
\newcommand{\slzbar}{\raise.15ex\hbox{$/$}\kern-.53em\hbox{$\bar{Z}$}}
\newcommand{\slQ}{\raise.15ex\hbox{$/$}\kern-.53em\hbox{$Q$}}
\newcommand{\slK}{\raise.15ex\hbox{$/$}\kern-.53em\hbox{$K$}}
\newcommand{\slkbar}{\raise.15ex\hbox{$/$}\kern-.53em\hbox{$\bar{k}$}}
\newcommand{\slkone}{\raise.15ex\hbox{$/$}\kern-.53em\hbox{$k_1$}}
\newcommand{\slpone}{\raise.15ex\hbox{$/$}\kern-.53em\hbox{$p_1$}}
\newcommand{\slpbarone}{\raise.15ex\hbox{$/$}\kern-.53em\hbox{$\bar{p}_1$}}
\newcommand{\slptwo}{\raise.15ex\hbox{$/$}\kern-.53em\hbox{$p_2$}}
\newcommand{\slpbartwo}{\raise.15ex\hbox{$/$}\kern-.53em\hbox{$\bar{p}_2$}}
\newcommand{\slqone}{\raise.15ex\hbox{$/$}\kern-.53em\hbox{$q_1$}}
\newcommand{\slD}{\raise.15ex\hbox{$/$}\kern-.53em\hbox{$\!D$}}
\newcommand{\slC}{\raise.15ex\hbox{$/$}\kern-.53em\hbox{$C$}}
\newcommand{\slA}{\raise.15ex\hbox{$/$}\kern-.73em\hbox{$A$}}
\newcommand{\slSigma}{\raise.15ex\hbox{$/$}\kern-.53em\hbox{$\Sigma$}}
\newcommand{\slpartial}{\raise.15ex\hbox{$/$}\kern-.53em\hbox{$\partial$}}
\newcommand{\slcalP}{\raise.15ex\hbox{$/$}\kern-.63em\hbox{$\cal P$}}
\newcommand{\sleps}{\raise.15ex\hbox{$/$}\kern-.53em\hbox{$\epsilon$}}
\newcommand{\slepsbar}{\raise.15ex\hbox{$/$}\kern-.53em\hbox{$\overline{\epsilon}$}}
\newcommand{\slepsstar}{\raise.15ex\hbox{$/$}\kern-.53em\hbox{$\epsilon$}^\star}
\newcommand{\slS}{\raise.15ex\hbox{$/$}\kern-.73em\hbox{$S$}}

\newcommand{\bb}{\mathbf}

\newcommand{\bk}{\mathbf{k}}
\newcommand{\bp}{\mathbf{p}}
\newcommand{\bq}{\mathbf{q}}

\newcommand{\bx}{\mathbf{x}}

\newcommand{\dtwo}[1]{\frac{\dd^2 #1}{(2\pi)^2}}

\newcommand{\p}{\prime}

\begin{document}
\title{Sudakov double logs in single-inclusive hadron production in DIS at small $x$ from the Color Glass Condensate formalism}


\author{Tolga Altinoluk$^{a}$, Jamal Jalilian-Marian$^{a,b,c,d}$ and Cyrille Marquet$^{e}$}

\affiliation{$^{a}$Theoretical Physics Division, National Centre for Nuclear Research,
Pasteura 7, Warsaw 02-093, Poland\\
$^{b}$ Department of Natural Sciences, Baruch College, CUNY, 17 Lexington Avenue, New York, NY 10010, USA\\
$^{c}$  City University of New York Graduate Center, 365 Fifth Avenue, New York, NY 10016, USA\\
$^{d}$ Theoretical Physics Department, CERN, 1211 Geneva 23, Switzerland\\
$^{e}$ CPHT, CNRS, \'Ecole polytechnique,  Institut Polytechnique de Paris, 91120 Palaiseau, France}


\begin{abstract}
We investigate the high $Q^2$ (photon virtuality) limit of single-inclusive hadron production
in DIS (SIDIS) at small $x$, using the color glass condensate formalism at next-to-leading order.
We focus on the $\Lambda_{QCD}^2 \ll \bp_h^2 \ll Q^2$ kinematic regime where $\bp_h$ is the produced
hadron transverse momentum, and extract the Sudakov double logarithms.
We further argue that compatibility between the CGC calculation and TMD factorization at one-loop
order can only be achieved if the small-x evolution is kinematically constrained.
\end{abstract}

\maketitle



\section{Introduction}

The Color Glass Condensate (CGC) effective theory \cite{Gelis:2010nm,Albacete:2014fwa,Blaizot:2016qgz} is commonly used to describe hadronic scattering processes at high collision energies in the Regge-Gribov limit. This effective theory is based on the gluon saturation phenomena which can be briefly described as follows; in the Regge-Gribov limit the increase in collision energy leads to a decrease of the longitudinal momentum fraction carried by the interacting partons. With decreasing $x$ the gluon density of the interacting hadrons increases rapidly. This rapid increase in the density is tamed by nonlinear interactions of the emitted gluons and cause the above mentioned gluon saturation phenomena at sufficiently high energies. 
The non-linear functional evolution equation with increasing energy (or equivalently with rapidity) is given by the Balitsky-Kovchegov~/~Jalilian-Marian-Iancu-McLerran-Wiegert-Leonidov-Kovner (BK-JIMWLK) equation \cite{Balitsky:1995ub,Kovchegov:1999yj,Kovchegov:1999ua,Jalilian-Marian:1996mkd,Jalilian-Marian:1997qno,Jalilian-Marian:1997jhx,Jalilian-Marian:1997ubg,Kovner:2000pt,Weigert:2000gi,Iancu:2000hn,Iancu:2001ad,Ferreiro:2001qy}.


Even though hints of gluon saturation phenomena have been seen in the experimental data from the Relativistic Heavy Ion Collider (RHIC) in the USA and the Large Hadron Collider (LHC) at CERN, a conclusive evidence is expected to be seen at the Electron Ion Collider (EIC) to be built in the USA. Deep inelastic scattering (DIS) on a dense target is one of the processes that will be at the focus of EIC to study the gluon saturation effects since it provides a clean environment to probe saturation. Theoretical computations of DIS related observables are frequently performed in the dipole factorization framework \cite{Bjorken:1970ah,Nikolaev:1990ja}, where the incoming lepton emits a virtual photon which splits into a quark-antiquark pair that scatters on the target. The splitting of the virtual photon into quark-antiquark pair is computed perturbatively while the interaction of the pair with the target is treated in the CGC framework by encoding the rescattering effects in the Wilson lines. 

With the advent of the EIC, there have been a lot of efforts to increase the precision of the theoretical calculations of DIS related observables. Inclusive DIS   \cite{Balitsky:2010ze,Balitsky:2012bs,Beuf:2011xd,Beuf:2016wdz,Beuf:2017bpd,Ducloue:2017ftk,Hanninen:2017ddy} and its fits to HERA data \cite{Beuf:2020dxl} for massless quarks have been computed at next-to-leading order (NLO) in strong coupling $\alpha_s$. Quark mass has been included in the NLO computations of inclusive DIS in  \cite{Beuf:2021qqa,Beuf:2021srj,Beuf:2022ndu} and fits of the results to HERA data have been performed in \cite{Hanninen:2022gje}. Single inclusive jet/hadron production in DIS \cite{Caucal:2024cdq,Bergabo:2024ivx,Bergabo:2022zhe} and Drell-Yan production \cite{Taels:2023czt} have been studied at NLO. Inclusive dijet (and/or dihadron) production in DIS have been computed at NLO in \cite{Caucal:2021ent,Taels:2022tza,Caucal:2022ulg,Caucal:2023nci,Caucal:2023fsf,Bergabo:2023wed,Bergabo:2022tcu,Iancu:2022gpw}. Finally, different aspects of diffraction and diffractive jet/dijet production have been studied in detail at NLO \cite{Beuf:2022kyp,Beuf:2024msh,Fucilla:2023mkl,Boussarie:2016ogo,Boussarie:2019ero,Fucilla:2022wcg}. 


A remarkable aspect of di-jet production is studied in \cite{Dominguez:2010xd,Dominguez:2011wm,Marquet:2016cgx} where the equivalence between the CGC and transverse momentum dependent distributions (TMDs) have been shown once the appropriate limits are applied. Namely, the high energy limit of the di-jet cross section computed in TMD factorization, and the correlation limit (when the two jets are produced back-to-back) of the di-jet cross section computed in the CGC framework. In addition, both unpolarized and linearly polarized gluon TMDs emerge in the CGC framework \cite{Metz:2011wb,Marquet:2017xwy}. The production of three-particle final states has also been considered in \cite{Ayala:2016lhd,Ayala:2017rmh,Altinoluk:2018uax,Altinoluk:2018byz,Altinoluk:2020qet}. The equivalence between the CGC and TMD factorization frameworks are extended beyond the correlation limit for di-jet production by re-summing the kinematic twist corrections in \cite{Kotko:2015ura,vanHameren:2016ftb,Fujii:2020bkl,Altinoluk:2021ygv,Altinoluk:2019wyu,Altinoluk:2019fui,Boussarie:2021ybe} and this new framework is referred to as small-$x$ improved TMD (ITMD) factorization; it interpolates between the dilute (BFKL) limit of the CGC and the TMD limit of the CGC. 

An important question of whether the equivalence between the CGC and TMD frameworks holds beyond LO for dijet production in DIS in the back-to-back limit has been addressed in \cite{Taels:2022tza,Caucal:2022ulg}. It was shown that in order to get the correct Sudakov double logarithm that was conjectured in \cite{Mueller:2012uf,Mueller:2013wwa}, one should adopt a kinematically constrained BK evolution \cite{Beuf:2014uia,Iancu:2015vea,Watanabe:2015tja} to properly re-sum the rapidity divergences that arise at NLO. More generally, the use of a kinematically-constrained non-linear small-x evolution is rapidly becoming unavoidable in CGC calculations, something which was realised long ago in the context of linear BFKL evolution \cite{Kwiecinski:1996td,Kwiecinski:1997ee}. In addition, combining low-$x$ and Sudakov resummation has been the subject of intensive research in various alternative approaches \cite{Hautmann:2008vd,Deak:2009xt,Jung:2010si,Sun:2011iw,Deak:2011ga,Dooling:2014kia,Zhou:2016tfe,Hentschinski:2016wya,Bury:2017jxo,Hentschinski:2017ayz,Boer:2017xpy,Xiao:2017yya,Stasto:2018rci,Zhou:2018lfq,Marquet:2019ltn,Zheng:2019zul,Blanco:2019qbm,vanHameren:2019ysa,vanHameren:2020rqt,Hentschinski:2021lsh,Nefedov:2021vvy,Boer:2022njw,Caucal:2024nsb,Caucal:2024bae}.

In this paper, we focus on the single inclusive hadron production in DIS. In section \ref{SIDIS_LO}, we start from the di-hadron production cross section in DIS, integrate over the antiquark to get the single inclusive hadron production cross section at LO and discuss the kinematic region where one can expect the emergence of the Sudakov double logarithms once the next-to-leading order corrections to the cross section are included in the analysis. In section \ref{SIDIS_NLO}, we include the next-to-leading order corrections to SIDIS cross section, identify the diagrams that will contribute to the Sudakov double logarithms in the appropriate kinematic region and discuss the divergences that appear. In section \ref{double_logs} we discuss how to extract these double logarithms. Finally, in section \ref{conc} we present a brief discussion of our results. 

\section{Single inclusive hadron production in DIS at leading order}
\label{SIDIS_LO}
The dominant channel for single-inclusive hadron production in Deep Inelastic Scattering at small $x$ is the production of a quark-antiquark pair, 
either of which can hadronize and be measured. This is a two-step process; first the virtual photon splits into a quark-antiquark pair which subsequently scatters 
on the target proton or nucleus. The cross section for this process can be written as \cite{Dominguez:2011wm}:
\bea
\frac{\dd \sigma^{\gamma^*A \to q\bar{q} X}}
{\dd^2 \bb{p}\, \dd^2 \bb{q} \, \dd y_1 \, \dd y_2} &=& 
\frac{ e^2 Q^2(z_1z_2)^2 N_c}{(2\pi)^7} \delta(1-z_1-z_2)
\int \dd^8 \bx \left[S_{122^\prime 1^\prime} - S_{12} - S_{1^\prime 2^\prime} + 1\right] \nonumber \\
&& e^{i\bb{p}\cdot\bb{x}_{1^\p1}} e^{i\bb{q}\cdot\bb{x}_{2^\p2}} 
\bigg[4z_1z_2K_0(|\bb{x}_{12}|Q_1)K_0(|\bb{x}_{1^\prime 2^\prime}|Q_1) + 
\nonumber \\
&&  
(z_1^2 + z_2^2) \,
\frac{ \bb{x}_{12}\cdot \bb{x}_{1^\prime 2^\prime}}{|\bb{x}_{12}| |\bb{x}_{1^\prime 2^\prime}|} \, 
K_1(|\bb{x}_{12}|Q_1)K_1(|\bb{x}_{1^\prime 2^\prime}|Q_1) 
\bigg]
\eea
where ($\bp , y_1$) and ($\bq , y_2$) are the transverse momentum 
and rapidity of the produced quark and antiquark respectively, and $Q^2$ is the virtuality of the incoming photon. The QED coupling $e^2$ should be understood as encompassing the various (massless) quark flavors: $e^2 = 4\pi\alpha_{em}\sum_f e_f^2$. The first (second) term inside the big square bracket corresponds to 
contribution of longitudinal (transverse) photons. Multiple scattering of the partons on the dense target are encoded in the dipole ($S_{ij}$) and quadrupole ($S_{ijkl}$) operators that are defined as 
\begin{align}
S_{ij}=\frac{1}{N_c}{\rm tr}\big\langle V_iV^\dagger_j\big\rangle\;\; , \; \; S_{ijkl}=\frac{1}{N_c}{\rm tr}\big\langle V_iV^\dagger_jV_kV^\dagger_l\big\rangle\
\end{align}
where the index $i$ corresponds to the transverse coordinate ${\bf x}_i$ and the Wilson lines $V_i$ are given in terms of the background field of the target $A^-$ as
\begin{align}
V_i={\hat{\cal P}}\;  {\rm exp}\Big( ig\int dx^+A^-(x^+,{\bf x}_i)\Big) \; . 
\end{align} 
%
\noindent Furthermore $\bx_1 (\bx_2 )$ is the transverse coordinate of the quark (antiquark) going through the target in the amplitude while the primed coordinates correspond to the same in the complex conjugate amplitude. We have defined $z_1 \equiv p^+/ l^+$ and $z_2 \equiv q^+/l^+$ as the longitudinal momentum fractions carried by the final state quark and antiquark, relative to the photon's longitudinal momentum $l^+$. The rapidity of a parton is related to its momentum fraction via $\dd y_i = \frac{\dd z_i}{z_i}$. We are also using the following definitions and short hand notations,
\begin{align}
Q_i = Q\sqrt{z_i(1-z_i)}, \,\,\,\,\,\, \bx_{ij} = \bx_i - \bx_j,\,\,\,\,\,\, \dd^8 \bx = \dd^2 \bx_1 \, \dd^2 \bx_2\, \dd^2 \bx_{1^\p} \, \dd^2 \bx_{2^\p}.
\end{align}
Finally, we shall denote the center-of-mass energy of the photon-proton (or photon-nucleus) collision $W^2\!=\!2\ l^+P^- -Q^2$ where $P^-$ is the target longitudinal momentum.

In order to get the single inclusive production cross section, we choose to integrate over the antiquark in the final state.
Denoting $\dd^6 \bx \equiv \dd^2 \bx_1 \, \dd^2 \bx_2\, \dd^2 \bx_{1^\p}$, we get:
\bea
\frac{\dd \sigma^{\gamma^*A \to q (\bp,y)X}}{\dd^2 \bp\, \dd y_1} &=&
\frac{ e^2 Q^2 N_c}{(2\pi)^5} \int \dd z_2 \delta (1-z_1-z_2)\, (z_1^2 \, z_2)\, 
\int \dd^6 \bx \left[S_{11^\p} - S_{12} - S_{1^\p2} + 1\right] 
e^{i\bp\cdot\bx_{1^\p1}} 
\nonumber \\
&&
\bigg\{
4z_1z_2 K_0(|\bx_{12}|Q_1) K_0(|\bx_{1^\p2}|Q_1) 
+ 
(z_1^2 + z_2^2) \,
\frac{\bx_{12}\cdot \bx_{1^\p2}} {|\bx_{12}||\bx_{1^\p2}|} \, 
K_1(|\bx_{12}|Q_1) K_1(|\bx_{1^\p2}|Q_1) 
\bigg\} .
\label{LOdsig-sidis-quark}
\eea
As hadronization of a colored parton is a genuinely non-perturbative phenomenon it 
is common to describe it by a parton-hadron fragmentation function when considering
hadron production at moderate to high transverse momenta. We will follow this approach 
here and convolute the partonic cross section with a quark-hadron fragmentation function
to get 
\bea
\frac{\dd \sigma^{\gamma^*A \to h (\bp_h,y_h)X}}{\dd^2 \bp_h\, \dd y_h} &=& 
\frac{ e^2 N_c}{(2\pi)^5} \int_{z_h}^1 \frac{dz_1}{z_h} D_{h/q}(z_h/z_1)\ z_1Q_1^2 \int \dd^6 \bx \left[S_{11^\p} - S_{12} - S_{1^\p2} + 1\right] 
e^{i(z_1/z_h) \bp_h\cdot\bx_{1^\p1}}
\nonumber \\
&&
\bigg\{ \frac{4Q_1^2}{Q^2} K_0(|\bx_{12}|Q_1) K_0(|\bx_{1^\p2}|Q_1) 
+ 
\Big[z_1^2 + (1-z_1)^2\Big] \,
\frac{\bx_{12}\cdot \bx_{1^\p2}} {|\bx_{12}||\bx_{1^\p2}|} \, 
K_1(|\bx_{12}|Q_1) K_1(|\bx_{1^\p2}|Q_1) 
\bigg\} 
\label{LOdsig-sidis-hadron}
\eea
where $z_h$ is fraction of the photon momentum carried by the produced hadron. We
note that in collinear fragmentation the produced hadron and (massless) parton rapidities 
are the same.  

In this work we are interested in particle production when the photon virtuality is large, so
that we need the large $Q^2$ limit of our expressions. However taking the large $Q^2$ limit
of the Bessel functions is a bit tricky; their argument depend on $Q^2$ through the combination
with momentum fraction $z$ which is integrated over so that the argument of Bessel functions
$z_1 (1 - z_1) Q^2$ can go to zero even at very large $Q^2$. Clearly taking this limit requires
some care; to accomplish this, we reformulate the procedure presented in \cite{Marquet:2009ca}
to extract the leading $1/Q^2$ behavior. We introduce a delta function of the form 
\beg
Q_1^{2n}\, K_{(0,1)}(|\bx_{12}|Q_1) \, K_{(0,1)}(|\bx_{1^\p2}|Q_1) = \int_0^{Q^2/4} d\bar{Q}^2\ (\bar{Q}^2)^n K_{(0,1)}(|\bx_{12}|\bar{Q}) K_{(0,1)}(|\bx_{1^\p2}|\bar{Q})\
\delta\left[\bar{Q}^2 - z_1(1-z_1)Q^2\right]
\label{LargeQ_T1}
\eeg
and insert it in the various hadron-level cross-sections. We
then do the $z_1$ integral using the delta function via 
\beg
\delta\left[\bar{Q}^2 - z_1(1-z_1)Q^2\right] = \frac{\delta(z_1-z_+)}{Q^2|1-2z_+|} +
\frac{\delta(z_1-z_-)}{Q^2|1-2z_-|}\quad\mbox{with}
\quad z_{\pm}=\frac12\left(1\pm\sqrt{1-4\bar{Q}^2/Q^2}\right) \, .
\label{LargeQ_T2}
\eeg
As we are interested in the high $Q^2$ limit, we keep only the leading power (in $Q^2$) terms, which comes from $z_1= z_+ \sim 1$., the so-called {\it aligned jet} configuration 
in which the quark that fragments into the measured hadron carries almost all of the photon longitudinal momentum.
For the LO cross section this gives (transverse and longitudinal cases respectively):
%
%
%
\bea
\left.\frac{\dd \sigma^{\gamma^*A \to h (\bp_h,y_h)X}}{\dd^2 \bp_h\, \dd y_h}\right|_{T,LP} &=&\frac{1}{Q^2}
\frac{e^2 N_c}{(2\pi)^5} \frac{D_{h/q}(z_h)}{z_h} 
\int \dd^6 \bx \left[S_{11^\p} - S_{12} - S_{1^\p2} + 1\right] 
e^{i(\bp_h/z_h)\cdot\bx_{1^\p1}}
\nonumber \\
&&
\frac{\bx_{12}\cdot \bx_{1^\p2}} {|\bx_{12}||\bx_{1^\p2}|} \, 
\int_0^{\infty} d\bar{Q}^2 \bar{Q}^2 \, K_1(|\bx_{12}|\bar{Q}) K_1(|\bx_{1^\p2}|\bar{Q})\ ,
\label{LOdsig-sidis-hadronT-LP}
\eea

\bea
\left.\frac{\dd \sigma^{\gamma^*A \to h (\bp_h,y_h)X}}{\dd^2 \bp_h\, \dd y_h}\right|_{L,LP} &=&\frac{1}{Q^2}
\frac{e^2 N_c}{(2\pi)^5} \frac{D_{h/q}(z_h)}{z_h} 
\int \dd^6 \bx \left[S_{11^\p} - S_{12} - S_{1^\p2} + 1\right] 
e^{i(\bp_h/z_h)\cdot\bx_{1^\p1}}
\nonumber \\
&&
\frac4{Q^2}\int_0^{\infty} d\bar{Q}^2 \bar{Q}^4 K_0(|\bx_{12}|\bar{Q}) K_0(|\bx_{1^\p2}|\bar{Q})\  .
\label{LOdsig-sidis-hadronL-LP}
\eea

Since the SIDIS production cross section via longitudinal photon is suppressed by a power of $1/Q^2$ in the high virtuality limit compared to the SIDIS production cross section via transverse photon, we restrict ourselves to the latter in this study. Moreover, one can rewrite the leading power expression of the transverse cross section in terms of the quark TMD distribution as \cite{Marquet:2009ca,Xiao:2017yya}:





\bea
\left.\frac{\dd \sigma^{\gamma^*A \to h (\bp_h,y_h)X}}{\dd^2 \bp_h\, \dd y_h}\right|_{T,LP} &=&\frac{\pi e^2}{Q^2} \frac{D_{h/q}(z_h)}{z_h} 
xq(x,\bp_h/z_h)
\label{LO_factorized_exp}
\eea
where the small-x quark TMD distribution $xq (x,\bp)$ is given by
\bea
xq (x,\bp)&=&
\frac{2N_c}{(2\pi)^6} 
\int \dd^6 \bx\ e^{-i\bp\cdot\bx_{11^\p}}\ \left[S_{11^\p} - S_{12} - S_{1^\p2} + 1\right] 
\frac{\bx_{12}\cdot \bx_{1^\p2}} {|\bx_{12}||\bx_{1^\p2}|} \, 
\int_0^{\infty} d\bar{Q}^2 \bar{Q}^2 \, K_1(|\bx_{12}|\bar{Q}) K_1(|\bx_{1^\p2}|\bar{Q})\ .
\label{eq:def_q_TMD-coordinate}
\eea
Here, the variable $x$ represents the longitudinal momentum fraction of the quark in the target wave function before it gets hit by the virtual photon. It is related to the final state quark momentum by $xP^-\!- Q^2/(2 l^+)\!=\!p^-$, that is $x\!=\!p^- /P^-\!+\!x_B$ with $x_B=Q^2/(Q^2+W^2)$. At large $Q^2$, because $z_1 \simeq 1$, the value of $x$ is fixed by kinematics: $x=x_B[1+(\mathbf{p}_h^2/z^2_h)/Q^2+{\cal O}(1/Q^2)]\gtrsim x_B$. In {\ref{eq:def_q_TMD-coordinate}}, the $x$ dependence of $xq(x)$ enters through the dipole amplitudes, or more precisely, through the rapidity scale choice at which they are to be evaluated, as will be discussed in the next section.

The writing of the CGC result (\ref{LOdsig-sidis-hadronT-LP}), for the leading-power SIDIS cross-section at small-x, in the factorized form (\ref{LO_factorized_exp}) is not arbitrary: Eq.~(\ref{eq:def_q_TMD-coordinate}) corresponds to the gluon-splitting contribution to the quark TMD, and the consistency with the operator definition of the quark TMD at small x was checked in \cite{Marquet:2009ca}. This correspondence is actually more evident in momentum space, where Eq.~(\ref{eq:def_q_TMD-coordinate}) can be rewritten as:
\bea
xq (x,\bp)&=& \int \dd^2 \bk_g
\left( \frac{N_c}{8\pi^4} \frac{\bk_g^2}{\alpha_s}  \int \dd^2 \bx_1\,  \dd^2 \bx_{1^\p}\ S_{11^\p}\, e^{-i \bk_g\cdot\bx_{11^\p}} \right)
\left( 
\frac{1}{4\pi^2} 
\frac{\alpha_s}{\bk_g^2}
\int_0^{\infty} \dd \bar{Q}^2 
\left|\frac{\bk_g-\bp}{\bar{Q}^2 + (\bk_g-\bp)^2} 
+\frac{\bp}{\bar{Q}^2 + \bp^2}\right|^2\right)\ .
\label{eq:def_q_TMD-momentum}
\eea
Inside the first parenthesis, we have the dipole gluon TMD $xG^{(2)}(x,\bk_g)$, which corresponds to an operator definition with one past-pointing and one future-pointing gauge staples, in the fundamental representation \cite{Dominguez:2010xd,Dominguez:2011wm}.

The function inside the second parenthesis is nothing but the $\xi$-integrated $\bk_g$-dependent splitting function $P_{qg}$ where the daughter quark (resp. antiquark) has transverse momentum $\bp$ (resp. $\bk_g\!-\!\bp$) and longitudinal momentum fraction (now in the "$-$" direction) $\xi$ (resp. $1\!-\!\xi$). In order to see this, let us recall that originally the integration variable $\bar{Q}^2$ quantifies the off-shellness of the intermediate $q\bar{q}$ pair, and appears in the energy denominator of the $\gamma^*\to q\bar{q}$ light-front wave function as $1/(\bp^2+\bar{Q}^2)$. Moving to the frame where it is the target gluon which splits into the $q\bar{q}$ pair before the photon hits the quark, we may identify that denominator with the $-p^\mu p_\mu=\bp^2+\xi (\bk_g\!-\!\bp)^2/(1\!-\!\xi)$ of the quark propagator. Therefore one is led to consider the change of variable $\bar{Q}^2= \xi (\bk_g\!-\!\bp)^2/(1\!-\!\xi)$, which leads to
\bea
\frac{1}{4\pi^2} \frac{\alpha_s}{\bk_g^2}
\int_0^{\infty} \dd \bar{Q}^2  \left|\frac{\bk_g-\bp}{\bar{Q}^2 + (\bk_g-\bp)^2} +\frac{\bp}{\bar{Q}^2 + \bp^2}\right|^2
&=&\frac{1}{4\pi^2} \frac{\alpha_s}{\bk_g^2}
\int_0^{1} \dd \xi  \left|\frac{\bk_g-\bp}{|\bk_g-\bp|} +\frac{|\bk_g-\bp|\ \bp}{(1-\xi)\bp^2+\xi(\bk_g-\bp)^2}\right|^2\nonumber\\
&=& \int_0^{1} \dd \xi\ P_{qg}(\xi,\bk_g,\bp)\ .
\eea
The function $P_{qg}$ introduced above coincides with the off-shell splitting function originally calculated in the Appendix C of \cite{Catani:1994sq}, and further discussed in the high-energy-factorization literature \cite{Ciafaloni:2005cg,Hautmann:2012sh}. To be precise, our $P_{qg}$ is equation (13) of \cite{Hautmann:2012sh} divided by $\pi$ (in \cite{Hautmann:2012sh} that 1/$\pi$ factor is instead included together with the measure $\dd^2 \bk_g$). The leading-power contribution in the CGC calculation incorporates the transverse momentum convolution, but features $xq(x)\!=\! xG^{(2)}(x) \int d\xi P_{qg}(\xi)$ and not $xq(x)\!=\! \int_x d\xi\, (x/\xi)G^{(2)}(x/\xi) P_{qg}(\xi)$ in the longitudinal direction, indicating that in $(x/\xi)G^{(2)}(x/\xi)$, $\xi$ is naturally set to unity, in accordance with the longitudinal momentum ordering at small x. For completeness, we also recall that the $\xi$ integral may be performed: 
\bea
\int_0^{1} \dd \xi\ P_{qg}(\xi,\bk_g,\bp)= \frac{1}{2\pi^2} \frac{\alpha_s}{\bk_g^2}
\left(1+\frac{\bp\cdot(\bk_g-\bp)}{\bp^2-(\bk_g-\bp)^2}\ln\frac{\bp^2}{(\bk_g-\bp)^2}\right)\ .
\eea
As was noticed in \cite{Marquet:2009ca}, the convolution (\ref{eq:def_q_TMD-momentum}) implies that the large transverse momentum tail of the quark TMD, i.e. $xq(x,\bp)\sim 1/\bp^2$, is controlled by the dynamics of the splitting and not by the high-$\bk_g$ tail of the gluon TMD. 

Going back to the coordinate space expression (\ref{eq:def_q_TMD-coordinate}), let us finally introduce the so-called b-space TMD defined by:
\bea
xq(x,\bp)= \int \frac{\dd^2 \bx_{11^\p}}{(2\pi)^2}\ e^{-i\bp\cdot\bx_{11^\p}}\ 
x\tilde{q}(x,\bx_{11^\p})\ .
\label{def_q_TMD}
\eea
Note that the b variable used in the TMD literature is the "dipole" size $\bx_{11'}$ here in our work, where the dipole is made up of the quark in the amplitude at transverse position $\bx_1$ and the quark in the conjugate amplitude at transverse position $\bx_{1'}$, and {\it should not be confused with the impact parameter} ${\bf b}_\perp=(\bx_1 + \bx_1^\p)/2$. Denoting $\dd^4 \bx \equiv \dd^2 {\bf b}_\perp \, \dd^2 \bx_2$, we write:
\bea
\label{unintegrated_dist}
 x\tilde{q}(x,\bx_{11^\p})= \frac{2N_c}{(2\pi)^4}\int \dd^4 \bx 
 \left[S_{11^\p} - S_{12} - S_{1^\p2} + 1\right] 
\frac{\bx_{12}\cdot \bx_{1^\p2}} {|\bx_{12}||\bx_{1^\p2}|} \, 
\int_0^{\infty} d\bar{Q}^2 \bar{Q}^2 \, K_1(|\bx_{12}|\bar{Q}) K_1(|\bx_{1^\p2}|\bar{Q})\ .
\eea

%


%

\section{Single inclusive hadron production in DIS at NLO}
\label{SIDIS_NLO}

\begin{figure}[H]
\centering 
\includegraphics[width=70mm]{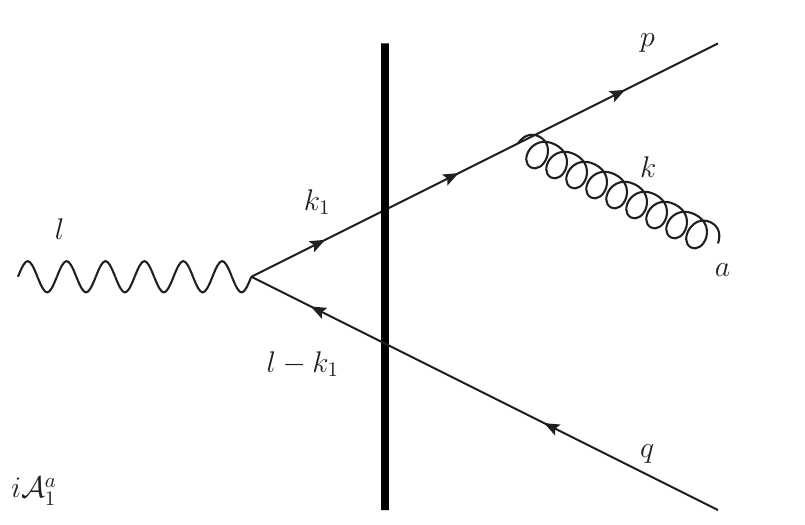}
\includegraphics[width=60mm]{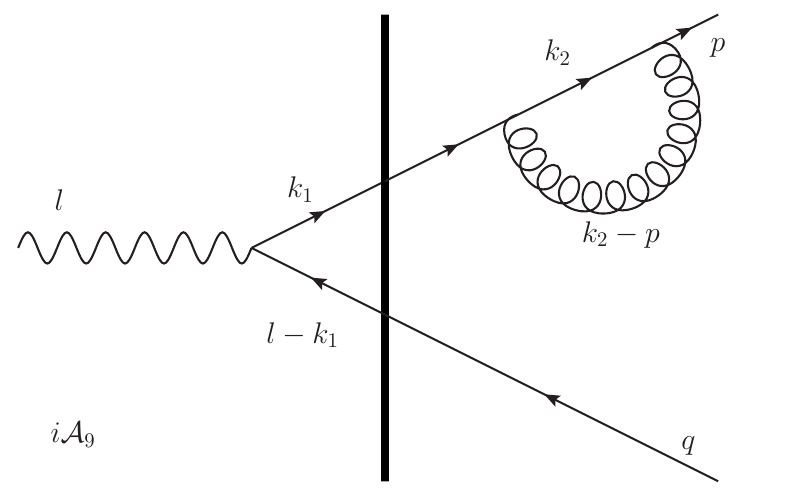}
\caption{The  NLO real (left) and virtual (right) diagrams giving large double logs. The arrows on fermion lines indicate fermion number flow, all momenta flow to the right, \textit{except} for gluon momenta. The thick solid line indicates interaction with the target.}\label{fig:nlodiags}
\end{figure}

Next to leading corrections to this leading order result have been calculated in~\cite{Bergabo:2022zhe}. In principle one must consider radiation of a gluon from either the quark or antiquark. This radiated gluon then can either be absorbed in the amplitude (virtual corrections) or in the complex conjugate amplitude (real corrections). In either case, the radiation can happen either after going through the target in which case only the original quark and antiquark scatter from the target, or before going through the target in which case all three partons scatter from the target. However as {\it we are interested only in terms in the the cross section which are enhanced by (Sudakov) 
double logs we will consider only the diagrams which contain a collinear divergence}. These are depicted in Fig. (\ref{fig:nlodiags}) where the labelling follows~\cite{Bergabo:2023wed,Bergabo:2022tcu}.

In this work we will focus on transversely polarized photons since the cross section 
with longitudinal photons is suppressed by $Q^2$ (photon virtuality) as compared with 
transverse photons. Partonic production cross sections are then obtained by squaring 
the production amplitude and including the appropriate phase space and flux factors. 
The contribution of the real correction is then given by~\cite{Bergabo:2022tcu} 

\begin{align}
&\frac{\dd \sigma_{1\times 1}^T}{\dd^2 \bp \, \dd^2 \bq\, \dd y_1 \, \dd y_2} = 
\frac{e^2 g^2 Q^2 z_2^2 (1-z_2) [z_1^2 z_2^2 + (z_1^2+z_2^2)(1-z_2)^2+ (1-z_2)^4]}
{(2\pi)^{10}z_1}
\int \frac{\dd z}{z} \delta(1-z-z_1-z_2) 
\nonumber \\ 
& \int \dd^{10} \bx \, K_1(|\bx_{12}|Q_2)K_1(|\bx_{1^\p 2^\p}|Q_2)\, 
\frac{\bx_{12}\cdot\bx_{1^\p 2^\p}}{|\bx_{12}||\bx_{1^\p 2^\p}|}\,
N_c  C_F 
[S_{122^\p 1^\p}-S_{12}-S_{1^\p 2^\p} +1] 
e^{i\bp\cdot\bx_{1^\p 1}} e^{i\bq\cdot\bx_{2^\p 2}}\,
\Delta^{(3)}_{1^\p 1} \,
e^{i\frac{z}{z_1}\bp\cdot\bx_{1^\p 1}} 
\end{align}
where we have defined the radiation kernel 
\begin{align}
\Delta^{(3)}_{ij} = \frac{\bx_{3i}\cdot\bx_{3j}}{\bx_{3i}^2 \bx_{3j}^2}
\end{align}
while the virtual correction is given by 
\begin{align}
\frac{\dd \sigma_9^{T}}{\dd^2 \bp \, \dd^2 \bq\, \dd y_1\, \dd y_2} = &
\frac{-e^2 g^2 Q^2 (z_1 z_2)^2(z_1^2+z_2^2)}{2(2\pi)^8} 
\int \dd^8 \bx \, K_1(|\bx_{12}|Q_1) K_1(|\bx_{1^\p 2^\p}|Q_1) \, 
\frac{\bx_{12}\cdot\bx_{1^\p 2^\p}}{|\bx_{12}||\bx_{1^\p 2^\p}|}  \, \delta(1-z_1-z_2)
\nonumber \\
&
N_c  C_F  
\big[S_{122^\p 1^\p} - S_{12} - S_{1^\p 2^\p} + 1\big]\, 
e^{i\bp\cdot\bx_{1^\p1}} e^{i\bq\cdot\bx_{2^\p2}}
\int_0^{z_1} \frac{\dd z}{z}\left[ \frac{z_1^2+(z_1-z)^2}{z_1^2}\right] 
\int \dtwo{\bk} \frac{1}{\left(\bk-\frac{z}{z_1}\bp\right)^2}\ .
\end{align}
Integrating over the antiquark phase space then gives the contribution of real and virtual 
corrections to single inclusive quark production:
\begin{align}
\frac{\dd \sigma_{1\times 1}^T}{\dd^2 \bp \, \dd y_1} = &
\frac{e^2 g^2 Q^2}{(2\pi)^{8}}\, 
\int_0^{1-z_1} \frac{\dd z}{z} \frac{(1-z-z_1) (z+z_1)}{z_1}  
\left[z_1^2 (1-z-z_1)^2 + \left(z_1^2 + (1-z-z_1)^2\right) (z+z_1)^2 + (z+z_1)^4\right]
\nonumber \\
&
\int \dd^{8}\bx \,  
K_1(|\bx_{12}|Q_{1z})K_1(|\bx_{1^\p 2}|Q_{1z}) 
N_c C_F \, [S_{11^\p} - S_{12} - S_{21^\p} +1] 
\frac{\bx_{12}\cdot\bx_{1^\p 2}}{|\bx_{12}||\bx_{1^\p 2}|}\,
\Delta^{(3)}_{11^\p}\,
e^{i\frac{z_1 + z}{z_1}\bp\cdot\bx_{1^\p 1}}\ .
\\
\frac{\dd \sigma_9^{T}}{\dd^2 \bp \, \dd y_1} = & 
- \frac{e^2 g^2 Q^2}{2(2\pi)^6} 
\int_0^{z_1} \frac{\dd z}{z}
(1-z_1) (z_1^2+(1-z_1)^2) \left[z_1^2 + (z_1-z)^2\right] 
\int \dd^6 \bx \, 
K_1(|\bx_{12}|Q_1) K_1(|\bx_{1^\p 2}|Q_1)
e^{i\bp\cdot\bx_{1^\p1}}
\nonumber \\
&
N_c C_F  \, [S_{11^\p} - S_{12} - S_{21^\p} + 1] 
\frac{\bx_{12}\cdot\bx_{1^\p 2}}{|\bx_{12}||\bx_{1^\p 2}|}
\int \dtwo{\bx_3} \frac{1}{\bx^2_{31}}\ ,
\end{align}
with
\begin{align}
Q^2_{1z} \equiv (1 - z - z_1) (z + z_1) Q^2\ .
\end{align}
Indeed, since we have integrated over $z_2$ the definition of $Q_1$ remains the same as before but $Q_2$ is now changed into $Q_{1z} $. For these NLO expressions, the extraction of the leading power in the large $Q^2$ limit can be achieved in a similar manner as in the previous section.

Let us start with the real-emission contribution. The hadron-level cross sections is
\begin{align}
\frac{\dd \sigma^{\gamma^*A \to h (\bp_h,y_h)X}_{{1\times 1}^T}}{\dd^2 \bp_h\, \dd y_h}
&= \frac{e^2 g^2}{(2\pi)^{8}}\, N_c C_F \, 
 \int_{z_h}^1 \frac{dz_1}{z_h} \ D_{h/q}(z_h/z_1)
\nonumber \\
& \, \times
\int_0^{1-z_1} \frac{\dd z}{z} \frac{Q_{1z}^2}{z_1} \left[z_1^2 (1-z-z_1)^2 + \left(z_1^2 + (1-z-z_1)^2\right) (z+z_1)^2 + (z+z_1)^4\right] 
\nonumber \\
& \times
\int \dd^{8}\bx [S_{11^\p} - S_{12} - S_{21^\p} +1] 
K_1(|\bx_{12}|Q_{1z})K_1(|\bx_{1^\p 2}|Q_{1z}) 
\frac{\bx_{12}\cdot\bx_{1^\p 2}}{|\bx_{12}||\bx_{1^\p 2}|}
e^{i\frac{z_1 + z}{z_1}\frac{z_1}{z_h}\bp_h\cdot\bx_{1^\p 1}}\Delta^{(3)}_{1^\p 1}\ , 
\label{eq:1x1-T}
\end{align}
and the leading-power contribution can be written as
\begin{align}
\left.\frac{\dd \sigma^{\gamma^*A \to h (\bp_h,y_h)X}_{{1\times 1}^T}}
{\dd^2 \bp_h\, \dd y_h}\right|_{LP}
&=\frac{\pi e^2}{Q^2}  \int \frac{\dd^2\bx_{11^\p}}{(2\pi)^2}\, e^{-i(\bp_h/z_h)\cdot\bx_{11^\p }} \, x\tilde{q}(x,\bx_{11^\p})
\nonumber \\
&
\times \frac{g^2C_F}{(2\pi)^{3}} \int_0^{1-z_h} \frac{\dd z}{z(1-z)}  \frac{D_{h/q}(z_h/(1-z))}{z_h} \left[1+(1-z)^2\right] 
\int \dd^{2}\bx_3 \, \Delta^{(3)}_{1^\p 1}\ , 
\end{align}
where we have used the definition of the b-space TMD (\ref{unintegrated_dist}).

Now let us discuss the virtual contribution. At hadron-level it reads 
\begin{align}
\frac{\dd \sigma^{\gamma^*A \to h (\bp_h,y_h)X}_{9^T}}{\dd^2 \bp_h\, \dd y_h}
&=- \frac{e^2 g^2}{2(2\pi)^6} N_c C_F 
 \int_{z_h}^1 \frac{dz_1}{z_1} \ \frac{D_{h/q}(z_h/z_1)}{z_h}
[z_1^2+(1-z_1)^2] \int_0^{z_1} \frac{\dd z}{z}\left[z_1^2 + (z_1-z)^2\right] \nonumber \\ 
&
\int \dd^6 \bx \frac{\bx_{12}\cdot\bx_{1^\p 2}}{|\bx_{12}||\bx_{1^\p 2}|}
Q_1^2 K_1(|\bx_{12}|Q_1) K_1(|\bx_{1^\p 2}|Q_1)
e^{i \frac{z_1}{z_h}\bp_h\cdot\bx_{1^\p1}}
\bigg[S_{11^\p} - S_{12} - S_{21^\p} + 1\bigg] \int \dtwo{\bx_3} \frac{1}{\bx_3^2}.
\label{eq:9-T}
\end{align}
Before extracting the leading-power, one should recall that in the full calculation, the UV divergence in this diagram is canceled by the UV divergent 
part of another virtual diagram, as explained in \cite{Bergabo:2022tcu} (labeled $\dd \sigma_{14(1)}$ there). We shall take this into account by introducing the scale $\mu$ in the $\bx_3$ integral. Then, the high-$Q^2$ limit is given by 
\begin{align}
\left.\frac{\dd \sigma^{\gamma^*A \to h (\bp_h,y_h)X}_{9^T}}
{\dd^2 \bp_h\, \dd y_h}\right|_{LP}
&
=\frac{\pi e^2}{Q^2}  \frac{D_{h/q}(z_h)}{z_h} 
\int \frac{\dd^2 \bx_{11^\p}}{(2\pi)^2} e^{-i(\bp_h/z_h)\cdot\bx_{11^\p}}  \, x\tilde{q}(x,\bx_{11^\p})
\nonumber \\ 
&
\times
\left(-
\frac{\alpha_sC_F}{4\pi^2}
\right)
\int_0^{1} \frac{\dd z}{z}\left[1 + (1-z)^2\right] \int_{|\bx_3|\mu>1}  \frac{\dd^2\bx_3}{\bx_3^2}.
\end{align}

Adding the LO and NLO (we need to multiply 
$\sigma^T_{9}$ by 2 since that contribution is the product of diagram $A_{9}$ in Fig.~\ref{fig:nlodiags} with the conjugate LO amplitude) terms we get
\begin{align}
\frac{\dd \sigma_{LO+NLO}^{\gamma^*A \to h (\bp_h,y_h)X}}
{\dd^2 \bp_h\, \dd y_h}\Bigg|_{LP}
=&
\frac{\pi e^2}{Q^2} \frac{1}{z_h}
\int \frac{\dd^2\bx_{11^\p}}{(2\pi)^2} \, e^{-i\frac{\bp_h}{z_h}\cdot\bx_{11^\p}} \, x\tilde{q}(x,\bx_{11^\p})
\bigg\{\left[1 -\frac{\alpha_sC_F}{2\pi^2}\int_0^{1} \frac{\dd z}{z}\left[1 + (1-z)^2\right] 
\int_{|\bx_3|\mu>1} \frac{\dd^2\bx_3}{\bx_3^2}\right] D_{h/q}(z_h) \,   
\nonumber 
\\
& \, + 
\left[
\frac{\alpha_sC_F}{2\pi^2} \int_0^{1 - z_h} \frac{\dd z}{z}  
\frac{\left[1 + (1 - z)^2\right]}{1-z} 
\int \dd^{2}\bx_3 \, \Delta^{(3)}_{1^\p 1}
\right]
D_{h/q}(\frac{z_h}{1 - z})
\bigg\}\ .
\label{LO+NLO_LP}
\end{align}

%

We next add (to virtual) and subtract (from the real part) the collinear divergence (for simplicity we also use $\mu$ as the factorization scale):


\begin{align}
\frac{\dd \sigma_{LO+NLO}^{\gamma^*A \to h (\bp_h,y_h)X}}
{\dd^2 \bp_h\, \dd y_h}\Bigg|_{LP}
=&
\frac{\pi e^2}{Q^2} \frac{1}{z_h}
\int \frac{\dd^2\bx_{11^\p}}{(2\pi)^2} \, e^{-i\frac{\bp_h}{z_h}\cdot\bx_{11^\p}} \, x\tilde{q}(x,\bx_{11^\p})
%
\nonumber 
\\
&
\hspace{-2cm}
 \times
\bigg\{D_{h/q}(z_h) + 
\left[\frac{\alpha_s C_F}{2\pi^2}
\left(
\int_0^{1 - z_h} \frac{\dd z}{z}\frac{1 + (1-z)^2}{1 - z} D_{h/q}(\frac{z_h}{1 - z})
- 
\int_0^1 \frac{\dd z}{z} [1 + (1-z)^2] D_{h/q}(z_h) \right)
\right]
\int_{|\bx_3|\mu>1} \frac{\dd^2\bx_3}{\bx_3^2}
\nonumber 
\\
&
+
\left[
\frac{\alpha_s C_F}{2\pi^2} 
\int_0^{1 - z_h} \frac{\dd z}{z}  
\frac{\left[1 + (1 - z)^2\right]}{1-z} 
\left(\int \dd^{2}\bx_3 \, \Delta^{(3)}_{1^\p 1} - \int_{|\bx_3|\mu>1} \frac{\dd^{2}\bx_3}{\bx_3^2}\right) 
\right]
D_{h/q}(\frac{z_h}{1 - z})
\bigg\}\ .
\label{eq:LO+NLO_LP}
\end{align}
%
The first two terms in the curly bracket in Eq.~(\ref{eq:LO+NLO_LP}) correspond to the leading-order cross section convoluted with DGLAP-evolved quark-hadron fragmentation function. Indeed, defined as the expectation value of two bare field operators which becomes singular in the short distance limit, the fragmentation function gets renormalized (evolves) by loop corrections where the renormalized quark-hadron fragmentation function is defined as (see \cite{Bergabo:2022tcu}) 

\begin{align}
D_{h/q}(z_h , \mu^2) \equiv 
\Bigg[ D^0_{h/q}(z_h) + \frac{\alpha_s C_F}{2\pi} 
\left( 
\int_0^{1 - z_h} \frac{\dd z}{z}\frac{1 + (1-z)^2}{1 - z} D_{h/q}(\frac{z_h}{1 - z})
- 
\int_0^1 \frac{\dd z}{z} [1 + (1-z)^2] D_{h/q}(z_h)  
\right)
\int_{1/\mu^2} \frac{\dd^2 \bx}{\bx^2} 
\Bigg]\ .
\label{evolved_FF}
\end{align}

Therefore, we can isolate the collinearly-divergent part of the cross section 
which leads to the standard DGLAP evolution of the fragmentation function (where the 
splitting function is defined with $+$ prescription) which is then convoluted with the 
Leading Order cross section. The result can be then be written as


\begin{align}
\frac{\dd \sigma_{LO+NLO}^{\gamma^*A \to h (\bp_h,y_h)X}}
{\dd^2 \bp_h\, \dd y_h}\Bigg|_{LP}
&=\;
d\sigma_{LO} \otimes D_{h/q}(z_h , \mu^2) + 
\frac{1}{Q^2} \frac{\pi e^2}{z_h}
\int \frac{\dd^2\bx_{11^\p}}{(2\pi)^2} \, e^{-i\frac{\bp_h}{z_h}\cdot\bx_{11^\p}} \, x\tilde{q}(x,\bx_{11^\p})
%
\nonumber \\
& \times\, 
\left[
\frac{\alpha_sC_F}{2\pi^2}
\left(\int \dd^{2}\bx_3 \, \Delta^{(3)}_{1^\p 1} - \int_{|\bx_3|\mu>1} \frac{\dd^{2}\bx_3}{\bx_3^2}\right) 
\int_0^{1 - z_h} \frac{\dd z}{z}  
\frac{\left[1 + (1 - z)^2\right]}{1-z} 
D_{h/q}(\frac{z_h}{1 - z})
\right]\ .
\label{eq:LO+NLO_LP-DGLAP-subtracted}
\end{align}
In the second line of that equation, one can already point out the origin of the Sudakov logarithms. They will come from the integration over 
$\bx_3$, in the range between and $1/\mu$ and $\bx_{11'}$ where the virtual (subtracted, to be precise) term does not get canceled by the real-emission one. Indeed, the combination inside the parenthesis is infrared finite, and in the real-emission integration, small dipole sizes are naturally cut-off by $x_{11'}$, while for the virtual term the integration over $\bx_3$ extends down to $1/\mu$. From now on, since we are only after double logs, we may set the scale as $\mu=Q$, which is the natural choice in SIDIS. We note however that scale setting is more intricate for the extraction of the single logarithms. The $\mu$ dependence of $D_{h/q}(z_h)$ may be applied in the $\alpha_s$ correction as well, since $ D_{h/q}(z_h,\mu^2) \simeq D_{h/q}(z_h) + O (\alpha_s)$ will generate NNLO corrections beyond the accuracy of our calculation.

The final step is to isolate the rapidity divergences which will result in 
the JIMWLK evolution of the dipole amplitude. To do so we introduce $z_f$, a factorization
scale in $z$ and break up the $z$ integral in Eqs. (\ref{eq:LO+NLO_LP-DGLAP-subtracted}) 
into two regions as follows,
\begin{align}
\int_0^{1 - z_h} \dd z = \int_0^{z_f} \dd z + \int_{z_f}^{1-z_h} \dd z
\label{zf_integral}
\end{align} 
where the first term contains the rapidity divergence while the second term 
is rapidity-finite and constitute part of the NLO corrections to the production
cross section. We may now write:


\begin{align}
\frac{\dd \sigma_{LO+NLO}^{\gamma^*A \to h (\bp_h,y_h)X}}
{\dd^2 \bp_h\, \dd y_h}\Bigg|_{LP}
& = 
d\sigma_{LO} \otimes D_{h/q}(z_h ,Q^2) + d\sigma_{NLO-rap-finite} 
\nonumber 
\\
&+
\frac{\pi e^2}{Q^2}\,  \frac{D_{h/q} (z_h,Q^2)}{z_h} 
\int \frac{\dd^2\bx_{11^\p}}{(2\pi)^2} \, e^{-i\frac{\bp_h}{z_h}\cdot\bx_{11^\p}} \, x\tilde{q}(x,\bx_{11^\p})\nonumber\\
&\times\left[
\frac{\alpha_s C_F}{\pi^2} 
\left(\int \dd^{2}\bx_3 \, \Delta^{(3)}_{1^\p 1} - \int_{|\bx_3|Q>1} \frac{\dd^{2}\bx_3}{\bx_3^2}\right) 
\int_0^{z_f} \frac{\dd z}{z}  
\right]\ .
\label{eq:lp-sing-coordinate}
\end{align}
It is customary in CGC calculations to absorb the entire second line of this expression into the leading-order term, making the dipole amplitudes $z_f$ dependent and evolving according to the Leading Log (LL) BK-JIMWLK equation. That was shown explicitly for the double-inclusive cross-section -- and consequently for SIDIS -- in \cite{Caucal:2021ent,Taels:2022tza,Bergabo:2022zhe}. Because the $z\to 0$ limit does not contribute to the evolution of the fragmentation function $D_{h/q}(z_h ,Q^2)$ (the collinear divergence in (\ref{evolved_FF}) cancels in that limit), one may safely send the integration limit $|\bx_3|>1/Q$ to zero in the rapidity subtraction term; the apparent UV divergence will cancel when the rapidity-subtraction terms from all the diagrams are put together to build the evolution equation.

In the rapidity-finite term however, that $|\bx_3|>1/Q$ lower limit must be kept, and this will generate a $\ln(|\bx_{|11^\p}|Q)$ factor. But this rapidity-subtraction procedure leaves us without double logarithms, in contradiction with TMD factorization. In the following section we will utilize a different subtraction scheme. In the NLO diagrams that possess only a rapidity divergence, that alternate scheme will generate single logs of $Q^2$ that would otherwise be absent. In the NLO diagrams that we considered in this section, which contain both a rapidity and a collinear divergence, the original single log will be turned into a double log, the very focus of the present work.

\section{Extraction of Sudakov logarithms}
\label{double_logs} 

The extraction of Sudakov double logs was addressed recently for the case of back-to-back di-jet production, where LL JIMWLK evolution of 
dipoles and quadrupoles resulted in Sudakov double logs with the wrong sign. To remedy this problem and restore compatibility with TMD factorization,
a kinematically-constrained JIMWLK evolution was introduced. We shall implement a similar idea in the present case of SIDIS at large $Q^2$.

The idea is that in addition to restricting the $+$ component of the gluon momentum $k^+ < z_f\ l^+$ with $z_f < 1$, i.e. $z < z_f$, we must also constrain their $-$ component $k^- > \tilde{z}_f\ |l^-|$ with $\tilde{z}_f > 1$ (we recall that $l^{\pm}$ refers to the momentum of the incoming virtual photon). This constraint enforces the gluon formation time $1/k^-=2k^+/\bk^2$ to be small enough to participate in the small-x evolution of the target: $1/k^- < (1/\tilde{z}_f)/|l^-|$ where $1/|l^-|= 2l^+ / Q^2$ represents the virtual photon lifetime. Gluons with formation time comparable to that of the photon lifetime must be excluded from the small-x evolution of the target and will contribute to the Sudakov phase space. A detailed discussion of how to implement such a constraint in the small-x evolution equations -- in a more general context than that of the resummation of Sudakov logarithms in two-scale processes -- can be found in the pioneering work \cite{Beuf:2014uia}. Here, we shall use a minimal subtraction scheme which allows to extract the double logarithms. 

Eventually, both $z_f$ and $\tilde{z}_f$ will have to be chosen in relation with the external kinematical variables of the process, and in order to insure the absence of additional large logarithms in the NLO finite pieces, the choices must respect $z_f \, \tilde{z}_f \sim 1$. The gluon lifetime constraint is then naturally written in momentum space as 
\begin{equation}
\Theta(\mbox{kin.const.}) = \Theta\left(z_f \frac{\bk^2}{Q^2} - z\right)
\end{equation}
and we will show that taking this into account in the rapidity subtraction terms, i.e. writing now
\begin{align}
\int_0^{1 - z_h} \dd z = \int_{z_f}^{1-z_h} \dd z + \int_0^{z_f} \dd z \left[1-\Theta(\mbox{kin.const.}) \right] + \int_0^{z_f} \dd z\ \Theta(\mbox{kin.const.}) 
\end{align} 
will restore the Sudakov double logarithms. Indeed, after absorbing a kinematically-constrained second-line of equation (\ref{eq:lp-sing-coordinate}) into the LO cross-section, in which the dipole amplitudes shall now evolve according to a kinematically-constrained JIMWLK equation, we are left with the following  $z$ integral:
\begin{equation}
\int_0^{z_f} \frac{\dd z}{z} 
\left[1-\Theta\left(z_f \frac{\bk^2}{Q^2} - z\right)\right] 
= 
\int_0^{z_f} \frac{\dd z}{z} 
\Theta\left(z - z_f \frac{\bk^2}{Q^2}\right) 
=
\Theta\left(Q^2-\bk^2\right) \ln\left(\frac{Q^2}{\bk^2}\right)\ .
\end{equation}
We then write our coordinate-space expression (\ref{eq:lp-sing-coordinate}) 
in momentum space using 
\begin{align}
\int \dd^{2}\bx_3 \, \Delta^{(3)}_{1^\p 1} - \int_{|\bx_3|Q>1} \frac{\dd^{2}\bx_3}{\bx_3^2}
=\int \, \frac{\dd^{2}\bk}{\bk^2}\ e^{i\bk\cdot \bx_{1^\p1}} - \int_{Q>|\bk|} \frac{\dd^{2}\bk}{\bk^2}
\end{align}
and we get 

\begin{align}
\frac{\dd \sigma_{LO+NLO}^{\gamma^*A \to h (\bp_h,y_h)X}}
{\dd^2 \bp_h\, \dd y_h}\Bigg|_{LP}
& = 
d\sigma_{LO}(z_f) \otimes D_{h/q}(z_h,Q^2) + d\sigma_{NLO-rap-finite}\nonumber\\
&+
\frac{\pi e^2}{Q^2}\,  \frac{D_{h/q} (z_h,Q^2)}{z_h} 
\int \frac{\dd^2\bx_{11^\p}}{(2\pi)^2} \, e^{-i\frac{\bp_h}{z_h}\cdot\bx_{11^\p}} \, 
x\tilde{q}(x,\bx_{11^\p})
\nonumber 
\\
&\times
\left\{
\frac{\alpha_s C_F}{\pi^2} 
\int^{Q^2} \, \frac{\dd^{2}\bk}{\bk^2} \left(e^{i\bk\cdot \bx_{1^\p1}} - 1\right)
 \ln\left(\frac{Q^2}{\bk^2}\right) 
\right\}\ .
%
\label{eq:lp-sing-coordinate-kin-constraint}
\end{align}
The leading order term corresponds to Eq~\ref{LO_factorized_exp}, but with the fragmentation function evaluated at the scale $\mu=Q$, and the quark TMD evaluated (though the dipole amplitudes) at the rapidity scale $\ln(1/z_f)$. Up to NNLO corrections, the $z_f$ dependence of $x\tilde{q}(x)$ may be applied in the $\alpha_s$ corrections as well.

The dipole amplitudes appear as $z_f$ dependent quantities because we have emitted the final-state gluon from the photon wavefunction. In practice however, the dipole amplitudes are obtained by implementing the small-x evolution from the target perspective, from an initial condition at some $x_0$ value down to a factorization variable in the "$-$" direction, chosen as the longitudinal momentum fraction $x$. Indeed, it is natural to work in a frame in which all the final-state particles other than the struck quark originate from partons in the target wavefunction with a longitudinal momentum fraction bigger than $x P^-$. This choice sets $\tilde{z}_f=x/x_B$ and then $z_f= x_B / x = Q^2/(\mathbf{p}_h^2/z^2_h+Q^2)\lesssim 1$, although again the value of $z_f$ is of no practical use when the evolution is done from the target perspective.

The final step is to perform the $\bp$ integral
\begin{equation}
\label{p_int}
\int^{Q^2} \frac{\dd^2 \bk}{\bk^2}  \left[ e^{- i \bk\cdot\bx_{11^\p}} -1 \right] \ln\left(\frac{Q^2}{\bk^2}\right)=
4\pi \int_0^{Q |\bx_{11^\p}|} \frac{\dd\tau}{\tau}  \left[ J_0(\tau)  - 1 \right] \ln\left(\frac{Q| \bx_{11^\p} |}{\tau}\right)
\end{equation}
which can be evaluated by using the following generic results: 
\begin{align}
\int_0^X\frac{d\tau}{\tau}\Big[J_0(\tau)-1\Big]=&-\ln\bigg(\frac{X}{c_0}\bigg)+O\bigg(\frac{1}{\sqrt{X}}\bigg)\\
\int_0^X\frac{d\tau}{\tau} \, \ln(\tau) \, \Big[J_0(\tau)-1\Big]=& -\frac{1}{2}\Big[ \ln(X)\Big]^2
+\frac{1}{2} \Big[\ln(c_0)\Big]^2+O\bigg(\frac{1}{\sqrt{X}}\bigg)
\end{align}
in the $X\to+\infty$ limit with $c_0=2\, e^{-{\gamma_E}}$. Using these results, 
the integral in Eq. \eqref{p_int} can be written as 
\begin{align}
4\pi \int_0^{Q |\bx_{11^\p}|} \frac{\dd\tau}{\tau}  \left[ J_0(\tau)  - 1 \right] 
\ln\left(\frac{Q| \bx_{11^\p} |}{\tau}\right)
=-\frac{\pi}{2}\ln^2\left(Q^2\bx_{11^\p}^2 / c_0^2\right) +O\bigg(\frac{1}
{\sqrt{Q |\bx_{11^\p}|}}\bigg)\ .
\end{align}
Then, we finally have "factorized" the contribution of the Sudakov double logs 
inside the "b-space" $\bx_{11^\p}$ integral.
Adding them to the LO cross-section gives the factor
\begin{equation}
1-\frac{\alpha_s C_F}{2\pi}\ln^2\left(Q^2\bx_{11^\p}^2 / c_0^2\right)=1-S_{sud}(\bx_{11^\p})\ .
\end{equation}
Further assuming exponentiation of the Sudakov logs we get the following result for
Sudakov re-summed single-inclusive hadron production in DIS,
%

\begin{align}
\frac{\dd \sigma_{LO+NLO}^{\gamma^*A \to h (\bp_h,y_h)X}}
{\dd^2 \bp_h\, \dd y_h}\Bigg|_{LP}
& = 
\frac{\pi e^2}{Q^2}\,  \frac{D_{h/q} (z_h,Q^2)}{z_h} 
\int \frac{\dd^2\bx_{11^\p}}{(2\pi)^2} \, e^{-i\frac{\bp_h}{z_h}\cdot\bx_{11^\p}} \, x\tilde{q}(x,\bx_{11^\p})\,  e^{-S_{sud}(\bx_{11^\p})} 
+ 
 d\sigma_{NLO-rap-finite}\ .
\label{eq:lp-sudakov}
\end{align}
This restore compatibility with TMD factorization \cite{Xiao:2017yya}. Note that we have obtained the global coefficient of the Sudakov double log, while in TMD factorization, half of that double log is assigned to the parton distribution function and the other half to the fragmentation function \cite{Ji:2004wu}. Hence our coefficient is twice the one indicated in e.g. \cite{Xiao:2017yya}, but our result is nevertheless consistent with TMD factorization.

\section{Conclusions and outlook}
\label{conc}

In this work we have studied the single-inclusive hadron production in DIS at NLO, focusing on the limit of large photon virtuality, i.e. large $Q^2$. In order to study this process, we started from the di-hadron production in DIS at NLO \cite{Bergabo:2022zhe,Bergabo:2023wed}, and integrated over the anti-quark to get the cross section for single-inclusive hadron production. As is known \cite{Marquet:2009ca}, at large $Q^2$ hadron production occurs predominantly via the so-called aligned-jet configuration, and processes initiated by longitudinally-polarized photons are suppressed by $1/Q^2$ when compared to production via transversely-polarized photons, thus we focused only on the transverse case. 

In the case of di-jet production at NLO, one can connect the CGC and the TMD factorization frameworks in the so-called correlation limit, where the two jets are produced almost back-to-back \cite{Taels:2022tza,Caucal:2022ulg}. In this kinematic region, the hard scale is set by the large transverse momentum of the produced jets while the semi-hard scale is set by the vector sum of the transverse momenta of the produced jets. In \cite{Taels:2022tza,Caucal:2022ulg}, it has been shown that in the case di-jet production in DIS, in order to get the correct sign of the large double logarithms of the ratio of hard to soft scales, which is known as the double Sudakov logarithms, one has to adopt kinematically constrained BK-JIMWLK evolution equation instead of the standard one when re-summing the rapidity divergences.

The extraction of the Sudakov double logarithms in the case of single inclusive hadron production in DIS is quite different from that in the case of back-to-back production. Indeed, when considering the SIDIS, the hard scale that appears in the argument of Sudakov logarithms is provided by the virtuality of the incoming photon $Q^2$, and the semi-hard scale by the hadron transverse momentum $\bp_h^2$. Imposing the ordering $\bp_h^2 \ll Q^2$ does not modify the Wilson line structures but rather acts on the photon splitting wave-function. The outcome of our study is that in this process, double Sudakov logarithms emerge only when the kinematically constraint BK-JIMWLK evolution equation is adopted for re-summing the rapidity divergences. 

In this work, we only focused on Sudakov double logarithms and a natural continuation would be to study the Sudakov single logarithms in SIDIS. Another interesting avenue of research to further connect the CGC and the TMD factorization frameworks in SIDIS will be to relax the eikonal approximation. Recently, studies were performed for inclusive DIS in \cite{Boussarie:2020fpb,Boussarie:2021wkn} and di-jet production in DIS in \cite{Altinoluk:2022jkk,Altinoluk:2023qfr,Agostini:2024xqs}. In SIDIS, including a $t$-channel quark exchange, which goes beyond the eikonal approximation, does appear necessary in order to complete the connection with the TMD picture, as in the eikonal case, the CGC only contains the gluon-splitting contribution to the quark TMD. These studies are left for future investigations.

\acknowledgements{We thank G. Beuf, R. Boussarie, P. Caucal, E. Iancu, P. Taels, B.W. Xiao and F. Yuan for useful discussions. TA and JJM thank CPHT, Ecole Polytechnique for hospitality during the visits when this work was initiated. JJM thanks CERN Department of Theoretical Physics for hospitality and financial support. 
This material is based upon work supported by the U.S. Department of Energy, Office of Science, Office of Nuclear Physics, within the framework of the Saturated Glue (SURGE) Topical Theory Collaboration.
JJM is supported by ULAM program of NAWA No. BPN/ULM/2023/1/00073/U/00001 and by the US DOE Office of Nuclear Physics through Grant No. DE-SC0002307.
This work has been performed in the framework of MSCA RISE 823947 ``Heavy ion collisions: collectivity and precision in saturation physics'' (HIEIC) and has received funding from the European Union's Horizon 2020 research and innovation programme under grant agreement No. 824093.}

\bibliography{mybib_New}
\bibliographystyle{apsrev}

\end{document}